%% file: template.tex
\documentclass{article}

\usepackage{arxiv}

\usepackage[utf8]{inputenc} 
\usepackage[T1]{fontenc}    
\usepackage{hyperref}       
\usepackage{url}            
\usepackage{booktabs}       
\usepackage{amsfonts}       
\usepackage{microtype}      
\usepackage{graphicx}

\usepackage[numbers]{natbib}
\usepackage{doi}
\usepackage{xspace}
\usepackage[most]{tcolorbox}
\usepackage{graphicx}
\usepackage{wasysym}

\usepackage{multirow}
\usepackage[normalem]{ulem}
\useunder{\uline}{\ul}{}

\renewcommand{\sp}[0]{security and privacy\xspace}

\renewcommand{\paragraph}[1]{
\vspace{2mm}
\noindent\textbf{#1}
}

\usepackage[noorphans,vskip=0.0pt,leftmargin=.1pt]{quoting}

\newtcolorbox{SummaryBox}[1]{enhanced,arc=1mm,outer arc=1mm,
  boxrule=0mm,toprule=0mm,bottomrule=0mm,left=1mm,right=1mm,leftrule=2pt,
  titlerule=0mm,toptitle=0mm,bottomtitle=0mm,top=0mm,
  colframe=blue!50!black,colback=blue!5!white,coltitle=blue!50!black,
  colbacktitle=yellow!50!white,colback=green!5!white,
  title=#1,
  fonttitle=\bfseries\sffamily\normalsize,fontupper=\normalsize\itshape,
}

\title{\textit{``It's not approved, but many, like myself, ignore the rule''}: Investigating the Landscape and Consequences of Unsanctioned Technology Use in Educational Institutes}

\author{ 
\textbf{Easton Kelso}\\
\texttt{eakelso@asu.edu}\\
Arizona State University
\and
\textbf{Ananta Soneji}\\
\texttt{asoneji@asu.edu}\\
Arizona State University
\and 
\textbf{Syed Zami-Ul-Haque Navid}\\
\texttt{snavid2@asu.edu}\\
Arizona State University
\and \\
\textbf{Yan Soshitaishvili}\\
\texttt{yans@asu.edu}\\
Arizona State University
\and\\
\textbf{Sazzadur Rahaman}\\
\texttt{sazz@cs.arizona.edu}\\
University of Arizona\\
\and\\
\textbf{Rakibul Hasan}\\
\texttt{rakibul.hasan@asu.edu}\\
Arizona State University
}

\renewcommand{\shorttitle}{Education Technology Acquisition Practices}

\hypersetup{
pdftitle={\textit{``It's not approved, but many, like myself, ignore the rule''}: Investigating the Landscape and Consequences of Unsanctioned Technology Use in Educational Institutes},
pdfauthor={Easton Kelso, Ananta Soneji, Rakibul Hasan},
pdfkeywords={First keyword, Second keyword, More},
}

\begin{document}
\maketitle

\begin{abstract}
Educators regularly use unsanctioned technologies---apps not formally approved by their institutions---for teaching, grading, and other academic tasks. 
While these tools often support instructional needs, they raise significant privacy, security, and regulatory compliance concerns. Despite its importance, understanding the adoptions and risks from the perspective of educators, who serve as de facto decision-makers behind unsanctioned technology use, is largely understudied in existing literature.

To address this gap, we conducted two surveys: one with $375$ educators who listed $1,373$ unsanctioned apps, and another with $21$ administrators who either often help educators to set up educational technologies (EdTechs) or observe their security or privacy incidents. Our study identified $494$ unique applications used by educators, primarily for pedagogical utility $(n=213)$ and functional convenience $(n=155)$, and the associated risks were often ignored. In fact, despite security and privacy concerns, many educators continued using the same apps ($n = 62$), citing a lack of alternatives or heavy dependence as barriers to discontinuation. We also found that fewer than a third of educators were aware of any institutional policy on \utech~use (K–12: $30.3\%$, HEI: $24.8\%$), and $22$ knowingly violated such policies. While $107$ received formal warnings, only $33$ adjusted their behavior. Finally, we conclude by discussing the implications of our findings and future recommendations to minimize the risks.

\end{abstract}


\input{1-intro}

\input{2-lit-review}

\input{3-educator-survey}
\input{3.1-survey-results}
\input{4-admin-survey}
\input{5-discussion}

\bibliographystyle{unsrt}
\bibliography{references,references-rakib, references-syed, online}  
\input{7-appendix}

\end{document}

%% file: 1-intro.tex
\section{Introduction}

%
%

The ubiquity of digital technologies in K-12 and higher education institutions (HEIs) introduces new security and privacy risks. For instance, the use of digital tools for classroom management, grading, and communications produces enormous digital footprints that often span beyond classrooms~\cite{datafication-HE, the_case_of_canvas}. 
Such datafication has positioned the education sector as a prime target for modern data breaches and unauthorized data sharing practices~\cite{education_sector_risk, us-schools-data-breach, pearson-data-breach, nyc-data-breach}. Although regulations such as the Family Educational Rights and Privacy Act (FERPA) and the Children’s Online Privacy Protection Act (COPPA) are designed to safeguard student and institutional data from misuse, their protections are often outpaced by the rapidly evolving technological landscape.

In response, educational institutions follow a structured procurement process for acquiring new technologies, involving steps such as vendor evaluation, contract negotiation, and compliance review to ensure alignment with institutional policies and regulatory standards. Kelso~\textit{et al.}~\cite{kelso_trust_2024} noted that even these \textit{sanctioned} technologies offer limited visibility to the educational institutes into their data protection and privacy practices. Nonetheless, the presence of formal contracts provides institutions with a means to hold vendors accountable as it serves as a ``guardrail'' in the event of security and privacy incidents. To make the situation worse, recently, a trend known as \textit{Shadow IT} has begun to surface in educational contexts where such guardrails are often non-existent~\cite{van_acken_poster_2023}. \textit{Shadow IT} refers to the use of software, hardware, or services without formal approval from institutional IT departments~\cite{silic_shadow_2014, walterbusch_missing_2017}. Its use has been linked to a range of risks, including data loss, weakened security postures, and compliance violations~\cite{acken_who_2024}.

While the use and associated risks of unsanctioned technologies (or “Shadow IT”) have been explored in corporate settings (\S~\ref{sec:use}), research on technology adoption in educational contexts has largely centered on institutionally acquired EdTech~\cite{student-perception-review, we-are-tracked, do-they-care, faculty-views-privacy, kelso_trust_2024, reyes2018won}. A few studies have shown that educators adopt technologies outside of institutional approval processes~\cite{cohney_virtual_2021, chanensonUncoveringPrivacySecurityK12CHI23}, but their main focus was to understand the risks from the institution's IT administrator perspective, overlooking the role of educators as everyday decision-makers who introduce unsanctioned tools into classrooms. 
We argue that, in this context, educators, as \textit{de facto} technology acquirers, make decisions that carry privacy, security, and legal consequences for students and institutions alike; thus, understanding their practices, threat perceptions, and the consequences of their decisions is critical to designing effective governance frameworks and institutional policies that can mitigate risks.

To address this gap, we study educators' acquisition practices of shadow IT through the perspectives of educators themselves.
We focus on how they independently select and implement tools for teaching-related activities (including applications, services, and devices) outside institutional approval channels, a practice we refer to as {\em Unsanctioned technology} use. 
Specifically, \textit{we conducted a sequential survey study involving both educators and administrators.}
First, we surveyed educators across K–12 and higher education institutes (HEIs) to investigate how unsanctioned technology is discovered, selected, and used, as well as to understand educators' awareness of data privacy and regulatory compliance.
We then conducted a follow-up survey with system administrators who work closely with educators to understand how institutions perceive and respond to the consequences of these informal adoptions.

Our study is guided by the following research questions:
\begin{enumerate}

\item \textbf{RQ1}: How are unsanctioned technologies discovered, adopted, and used by educators in K–12 and HEIs? 

\item \textbf{RQ2}: What are the perceived and experienced security and privacy implications with~\utech~use?

\item \textbf{RQ3}: How do institutional and personal factors influence~\utech~use?

\end{enumerate}

\noindent
By exploring the above questions, we provide insights into the landscape and consequences of~\utech~use within K-12 and HEIs and highlight our study's key findings here:

\smallskip
\noindent\textbf{Educators use hundreds of~\utech}.
We surveyed $375$ educators who listed a total of $494$ unique apps, with tools such as  Google Classroom, Kahoot!, Canva, and ChatGPT among the most cited (\S~\ref{edtech_use_prevalence}). 
Educators tend to use multiple apps: in total, $1,373$ apps were cited, for an average of over 3 apps per educator.
Unsanctioned technology discovery (\S~\ref{edtech_discovery_mechanism}) was driven by informal channels---$290$ educators learned about tools from colleagues. These apps and services were primarily adopted for \emph{pedagogical utility} ($n = 213$) and \emph{functional convenience} ($n = 155$) (\S~\ref{edtech_selection_criteria})---while continued use was often justified by ease of use ($n = 294$) and student engagement ($n = 108$).

\smallskip \noindent \textbf{Security \& privacy is the least cited criteria for acquiring~\utech.}
Notably, only $24$ educators mentioned \emph{security or privacy} as selection criteria, suggesting that data protection was not a primary consideration for most educators.
Additionally, $83\%$ educators reported using personal devices for teaching, with $231$ storing student data on them---$127$ of which had personal cloud backups enabled, raising further privacy risks.

\smallskip
\noindent\textbf{Unsanctioned technology use keeps educational institutes in the dark and severely undermines their security posture.} 
167 educators indicated that they encountered privacy or security-invasive behaviors (excessive data collection, third-party data sharing, and user tracking) that eventually led to the uninstallation of apps (\S~\ref{observed_risks}). 
Despite such concerns, many educators continued using the same apps ($n = 62$), citing lack of alternatives or deep integration into their workflow as barriers to discontinuation. Among those who did uninstall tools, only nine used a data deletion feature—suggesting a broader lack of awareness or availability of secure off-boarding mechanisms. 

To understand institutional perspectives, we surveyed $21$ administrators who closely worked with educators.
They listed a total of $16$ serious concerns posed by the use of \utech (\S~\ref{sec:it_observed}), with \emph{data leakage} ($n = 13$) being the most cited risk. 
These risks are compounded when \utech require students to register with institutional credentials, exposing institutional data to potential violations of federal data protection regulations (\S~\ref{sec:privacy:compliance}). 


\smallskip
\noindent\textbf{Institutional policy for \utech~use lack enforcement, while educators report strong security habits---but often misunderstand core privacy concepts.} 
Fewer than a third of educators were aware of any institutional policy on \utech~use (K–12: $30.3\%$, HEI: $24.8\%$), and $22$ knowingly violated them (\S~\ref{institute_policy_edtech_use}). 
While $107$ received formal warnings, only $33$ adjusted their behavior.
Despite this, enforcement was minimal---administrators described policies as ``rarely enforced'' or ``outdated,'' and seven even recommended \utech~themselves, citing resource constraints and governance gaps as barriers to proactive control over \utech~use  (\S~\ref{institute_culture_edtech_use}).
$96.8\%$ of participants reported strong personal security and privacy behavior ($\mu \geq $ 3.5 on SeBIS~\cite{Egelman2015ScalingSeBIS}) and yet, over $40\%$ of educators misunderstood basic privacy terms such as ``data broker'' or ``personally identifiable information (PII),'' exposing a critical gap between self-perceived competence and actual privacy literacy (\S~\ref{educator_sp_attitudes}).

Our contributions are summarized as follows.

\begin{enumerate}

\item We conduct the first study to surface the prevalence, factors, and consequences of \utech{} use in an educational context from the perspective of educators who play a central role in acquiring and using \utech{}. Specifically, we conducted a sequential survey study involving both educators and administrators who work closely with educators.

\item We report the empirical evidence of unsanctioned technology use in the educational (K-12 and HEI) context.  375 educators who took our survey reported a total of 494 unique \utech{}. We also find that while educators often discover these tools through peers, the most commonly cited reasons for using \utech{} are their pedagogical utility and functional convenience.

\item We highlight that, among many contributing factors, lack of policy awareness and enforcement, as well as misunderstandings around security and privacy risks, play a pivotal role in the continued use of \utech{} despite observing security and privacy consequences.

\end{enumerate}

%% file: 2-lit-review.tex
\section{Background and Related Work}

In this section, we review prior research on institutional adoption processes for educational technologies, the use of unsanctioned technology in both business and educational settings, and the relevant regulatory frameworks that shape technology use in U.S. education.

\subsection{Institutional adoption of technologies}
Digital technologies, including learning management systems, student monitoring apps, collaboration tools, and many others have become ubiquitous in all levels of education institutes in the US~\cite{kelso_trust_2024, chanensonUncoveringPrivacySecurityK12CHI23}. When institutionally acquired, they go through security and privacy evaluations before deployment, as well as typically monitored by institutional security experts while in use~\cite{kelso_trust_2024}. While these tools have security and privacy implications, as revealed by many past studies~\cite{cohney_virtual_2021, chanensonUncoveringPrivacySecurityK12CHI23, balash2023educators}, institutional acquisition is preceded by legal contracts with the vendors that specify what data can be collected and how they can be used~\cite{kelso_trust_2024}. This sanctioning process, comprising security evaluations and contract negotiations, provides a layer of protection for students' data from being misused. In contrast, when technology adoption bypasses such a sanctioning process, it might impose greater privacy, security, and compliance threats to both the users and institutions. While a few prior studies briefly mentioned unsanctioned technology use in education~\cite{cohney_virtual_2021, chanensonUncoveringPrivacySecurityK12CHI23}, their prevalence and impact have not been comprehensively investigated.

\subsection{Use of unsanctioned technology in business and education settings}\label{sec:use}
Several studies have investigated unsanctioned technology use in corporate settings (termed ``Shadow IT'' in this context~\cite{acken_who_2024}). Acken \textit{et al.} investigated the landscape of \emph{Shadow IT} usage in a corporate environment \cite{acken_who_2024}. They define Shadow IT as ``any software, hardware, and services that have been acquired without the knowledge of the IT department.'' Their findings include the prevalence of such use, limited awareness of company policy, and continued use despite perceived risks. 
In a similar work, Silic \textit{et al.} categorized unauthorized software, and identified the widespread use of applications that enhance productivity, and foster ease of communication and collaboration \cite{silic_shadow_2014}. They also found distrust of IT employees and lack of cybersecurity knowledge to be the driving forces behind the adoption of such tools. While the authors called for better policies, they acknowledged the shortcomings of additional restrictions. Haag \textit{et al.} showed that workplace culture, influence of colleagues, and acceptance from even non-users shape the motivation to use Shadow IT~\cite{haag_acceptance_2019}. 

Investigating unsanctioned technology in the education context is virtually non-existent. In a recent short article, Acken \textit{et al.} listed \utech, along with associated risks (such as lack of access control and visibility to threats), commonly found in Dutch higher education institutes based on interviewing 11 IT professionals. This study and studies on corporate settings can provide a glimpse of what we might expect from a similar investigation focusing on K-12 and HEIS in the US. However, the regulatory landscape in the US is significantly different than in the Netherlands (see \S~\ref{sec:regulations}), which greatly impacts data privacy protections and legal contracts~\cite{kelso_trust_2024}. We also expect to see differences, in terms of both technology use and associated risks, between K-12 and HEI settings because of differences in their structures, autonomous status, and needs. Both settings are also expected to hugely differ from the corporate settings; indeed, a recent study reported that privacy and compliance risks are much higher in educational institutes due to their open and collaborative environments compared to corporate entities~\cite{kelso_trust_2024}. Moreover, education institutes are not only consumers, but also producers of technologies, and cultivate a more exploration-friendly environment geared toward innovation---all of these might contribute to their adoption and use of technologies without institutional guardrails. Thus, investigating the \utech landscape and its security and privacy impacts in this setting is critical.

\subsection{Relevant federal regulations}\label{sec:regulations}
Unfortunately, there is no comprehensive federal privacy law in the US. There are, however, sector-specific laws that can apply to technology that access students' data. The major federal law is FERPA (Family Educational Rights and Privacy Act, which covers Personally Identifiable Information (PII) of students and other educational records {ferpa-faq}. The COPPA (Children's Online Privacy Protection Act) protects information about kids under 13~\cite{coppa}, thus any technology serving at K-12 schools most likely should comply with it. Finally, HIPAA (Health Insurance Portability and Accountability Act) might apply to tools that access health information~\cite{hipaa}.

Unfortunately, these laws have been criticized for being limited, vague, and containing loopholes that are regularly being exploited to collect, use, and sell consumers' private data~\cite{russell2018transparencystudentdata, zeide_learner_2018, brinkman_analysis_2013, paris2022sins, reidenberg_privacy_2013, skowronski2022coppa}. Since \utech is not restricted by institutional contracts, in the absence of comprehensive regulatory protections, privacy, security, and safety risks from them are likely to be severe and require urgent investigation.

%% file: 3-educator-survey.tex
\section{Methods}

We utilized survey methodology to unveil the landscape of~\utech~use at educational institutions in the U.S., security and privacy compliance issues with such a prevalent and invisible use of~\utech, and lastly, factors influencing~\utech~use.
Specifically, we employed a \textit{sequential survey design}~\cite{creswell2017designing}, which involved two online surveys---first, with educators (i.e., K-12 and HEI instructors) and later with system administrators from educational institutes---to gather experiences and perceptions from different perspectives on the use and impact of~\utech~use. Additional details about the sequential survey approach can be found in Appendix~\ref{why_sequential_design}.
Next, we present our methodology in details.

\subsection{Educator survey}
\smallskip
\noindent\textbf{\textit{Survey design.}} 
The survey---deployed on Qualtrics\footnote{\url{https://www.qualtrics.com/}}---was designed to collect both quantitative and qualitative insights into the use of unsanctioned technologies in education. 
It combined multiple-choice, Likert-scale, and open-ended questions to examine application usage, institutional awareness, security and privacy concerns, and data management practices.
The survey comprised four primary sections: (1) app-specific behaviors and decision-making, (2) personal device use and data handling practices, (3) institutional policy and behavior changes due to apps' \sp, and (4) digital privacy literacy and \sp behavior. 

\textit{In the first section,} we start by asking the participants to list up to five unsanctioned applications (mobile, web, or desktop) they had personally used for teaching, grading, or research. 
For each app listed, a structured series of follow-up questions were triggered to capture motivations for selection, discovery sources, integration with institutional tools, requirements of student registration with institutional credentials, current use status, and regulatory compliance awareness (e.g., ``Did you integrate [App Name] with institutionally licensed tools such as Canvas or Zoom?'', ``Was [App Name] compliant with FERPA or HIPAA?''). 
Participants were also asked about data migration, deletion options, and whether they had observed or reported privacy issues associated with each tool.

\textit{In the second section,} we asked questions related to the use of personal devices for professional tasks, including downloading student data to personal laptops or phones, and whether such devices had automatic cloud backups enabled. 

\textit{In the third section,} we explored participants’ understanding of their institution’s policies on~\utech~use and related institutional communications. We also asked whether any security or privacy concerns had influenced changes in their past general app usage behavior.

\textit{In the final section}, we included privacy literacy items, such as open-ended definitions of online tracking, data brokers, and personally identifiable information (PII), which were later coded for accuracy. Likert-scale questions measured participants' attitudes toward protecting privacy, trust in software, and security habits. 

Lastly, demographic questions concluded the survey. A hidden instruction (append ``No
more comments'' at the end) was included in the last question to identify and flag AI-generated submissions. The full survey instrument is provided in Appendix~\ref{educator_questionnaire}.

\noindent\textbf{\textit{Recruitment.}}
We recruited K-12 and HEI educators through Prolific~\cite{prolific2} as crowdsourcing platforms are effective tools for studying large and diverse groups of people in a short amount of time~\cite{hansson2019capitalizing, renard2019social}. The survey was made available to only current or former educators from K-12 and HEI institutions who met our inclusion criteria, such as experience using unsanctioned educational technologies. We piloted the survey with 50 participants and based on the responses improved the questions before collecting more data. Each participant received $\$5$ compensation for completing the survey, the median completion time was 15 minutes.


\subsection{Administrator survey}
\label{admin_survey}

\smallskip
\noindent\textbf{\textit{Survey design.}}
The administrator survey was developed to capture qualitative insights into institutional practices, oversight mechanisms, and policy responses regarding the use of unsanctioned technologies in educational settings. While primarily qualitative, the instrument also included structured multiple-choice questions to contextualize the policy environment and decision-making processes within institutions.
The survey covered three main sections: (1) experiences with technology integration, (2) institutional security/privacy practices and concerns, and (3) institutional policy-related communications and enforcement.

\textit{In the first section,} participants were asked whether they had ever recommended non-institutionally licensed tools to instructors and whether they were aware of faculty requests to integrate such tools with official platforms (e.g., Canvas or Google Classroom).
Subsequent questions explored how such integration requests are handled, their frequency, and whether they undergo security audits.

\textit{In the second section,} we addressed the \sp implications with~\utech~use.
Specifically, we asked who is responsible for responding to incidents involving personally acquired technologies and elicited participants’ perspectives on the most troubling risks posed by unlicensed tools.
Participants were also asked to describe any security or privacy incidents they had encountered. 
We ended this section by asking questions about barriers to proactive remediation of the implications from~\utech~use.

\textit{In the third section,} we addressed institutional policies and their awareness among educators concerning the use of non-licensed tools, downloading student data to personal devices, and how educators are informed about relevant policies.
\textit{Finally,} the survey ended with requesting for demographic information, job title, and institution type. 
A hidden instruction (append ``I am done'' at the end) was included in the last question to identify and flag AI-generated submissions. The full survey instrument is available in Appendix~\ref{admin_questionnaire}.

\noindent\textbf{\textit{Recruitment.}}
System administrators were recruited in a different approach than K-12 and HEI educators, as large-scale recruitment from this target group was not feasible through online survey platforms such as Prolific.
Due to the lack of a direct and centralized recruitment method, we employed a targeted outreach strategy, which included direct email invitations sourced from publicly available online resources and author contacts.
Additionally, the study link was shared on the EDUCAUSE platform\footnote{\url{https://connect.educause.edu/}} and within two relevant Reddit communities: r/k12sysadmin\footnote{\url{https://www.reddit.com/r/k12sysadmin/}} and r/k12cybersecurity\footnote{\url{https://www.reddit.com/r/k12cybersecurity/}}, to reach professionals engaged in educational IT administration and cybersecurity.

Participants were compensated through an opt-in lottery for a chance to win a $\$20$ gift card, a higher incentive compared to the educator survey due to the specialized nature of their expertise and the challenges in recruiting system administrators~\cite{acar2017security}. 
The survey was administered between December 2024 and January 2025.


\subsection{Data analysis}
\label{data_analysis}

Survey responses were first screened for quality and completeness. 
For both surveys, we manually identified and excluded AI-generated responses using a hidden prompt embedded in the final open-ended question and by reviewing for unnatural phrasing or third-person narration. Only fully completed and unique responses were retained for analysis.
For the \textbf{\textit{educator survey}}, quantitative data---comprising multiple-choice, and Likert-scale items---was analyzed using descriptive statistics (e.g., counts, percentages) to identify patterns in app integration behaviors, data handling practices, and privacy/security attitudes. 
Open-ended responses, including reasons, concerns, and privacy perceptions, were thematically coded~\cite{braun2022conceptual} by two researchers to extract recurring motivations, concerns, and behaviors. 
Coding was also done to identify the accuracy of short textual definitions of privacy-related terms. 
Any disagreements among coders were resolved through discussions.
For the \textbf{\textit{administrator survey}}, due to the smaller sample size and emphasis on open-ended questions, we conducted a thematic analysis~\cite{braun2022conceptual}. 
Two researchers independently reviewed responses to identify institutional practices, perceived risks, and gaps in policy implementation. 
Multiple-choice responses were used to contextualize trends and support interpretation of the qualitative insights.

\smallskip
\noindent\textbf{\textit{Privacy policy analysis.}}
To complement educators' self-reported data on their listed apps' compliance perception, we conducted a structured review of the privacy policies of twenty learning-related applications frequently mentioned by educators. 
Policies were manually annotated by three researchers to determine whether they claimed compliance with FERPA, COPPA, and HIPAA (when applicable), whether they permitted use by children under $13$, and whether they disclosed practices around data sharing, selling, and tracking. 
This review provided external validation of participants' privacy perceptions and helped identify mismatches between educator understanding and actual tool practices.


\subsection{Ethics}
While our study received exempt status by our Institutional Review Board (IRB), we adhered rigorously to ethical and privacy standards recommended for human subjects’ studies. 
This included getting informed consent for participation, anonymizing participants' personally identifiable information that participants gave when answering the survey, and ensuring that all collected data was stored securely and only used for research purposes.

\subsection{Study limitations}
\textit{First}, recruiting through Prolific might have introduced bias in our data, as participants were comfortable using online platforms and thus more tech-savvy than the general population of educators.


%
\textit{Second,} interpreting open-ended responses may introduce subjectivity, a common challenge in qualitative research. To reduce this risk, we engaged in collaborative theme refinement, conducted double-coding for consistency, and resolved disagreements through discussion. In addition, the depth of responses in open-ended questions varied among participants, i.e., some participants provided rich qualitative detail, while others gave brief, low-effort answers. We addressed this by filtering out brief and low-effort responses. However, the large sample size of educators who participated in our study allowed us to capture broad trends and key themes, providing a comprehensive understanding of the landscape.

\textit{Third,} some questions required participants to recall prior decisions or experiences, which may have introduced recall bias. To minimize ambiguity and improve the accuracy of recall, we phrased questions clearly and encouraged concrete examples.


\textit{Lastly,} as previously discussed~\ref{admin_survey}, system administrators in educational institutions---particularly those in K-12 settings---were challenging to recruit due to the limited availability of publicly accessible contact information. 
To maximize recruitment, we employed multiple outreach strategies, including advertising the survey on Reddit and EDUCAUSE platform, as well as offering higher compensation through an opt-in lottery. 
Despite these efforts, the final sample size for administrators remained relatively small. 
However, we ensured data quality and relevance by screening responses for completeness and authenticity, allowing us to capture meaningful insights from this group of participants.


Despite these limitations, this study provides a strong empirical foundation for understanding \utech~use.
However, findings may not fully generalize to all educational institutions, particularly those in regions or contexts not well represented in our sample.
Future research should explore these dynamics in broader and more diverse populations to validate and extend our results.

%% file: 3.1-survey-results.tex
\section{Unsanctioned Technology Landscape (RQ1)}
\label{rq1_results}

This section focuses on data from the educator survey. Based on both quantitative and qualitative responses, it portrays the landscape of \utech
 use in K-12 and higher education institute (HEI) settings. 


\subsection{Prevalence and categories}
\label{edtech_use_prevalence}

To map the landscape of~\utech, participants were first asked to \emph{list three to five (max) applications (apps) they currently use that had not been formally approved by their institutions}.
We also asked them if and how they use personal devices in teaching activities. We report results about apps followed by device use.

\subsubsection{Unsanctioned applications.}
In total, participants listed $1,373$ unsanctioned apps and services, of which $494$ were unique, highlighting their rampant use within the education ecosystem. It should be noted that institutions have different sanctioned technologies, meaning that a person's unsanctioned technology could be sanctioned at another institution.
A majority $374$ of the participants listed three apps, $161$ listed four applications, and $88$ listed five.
Among the $494$ unique apps, $88$ applications were shared between K-12 and HEI respondents. 
K-12 educators combined listed $284$ unique tools (with the top three being Google Classroom, Kahoot!, and Canva), while HEI educators listed $121$ unique tools (top three: ChatGPT, Kahoot!, and Canvas), highlighting both shared trends and context-specific needs across sectors. 
Table~\ref{tab:top_apps} presents the most commonly cited apps along with their respective categories. 

\begin{SummaryBox}{Takeaway \#1:}
A majority of educators in our study reported using three \utech platforms.
\vspace{-2mm}
\end{SummaryBox}

Apps were categorized mainly on two fronts: 1) \emph{pedagogical} (e.g., tools for instruction, assessment, or content creation) and \emph{functional convenience} (e.g., tools for communication, content delivery, or storage).
\textit{Alarmingly, some of the apps reported by educators were no longer supported by their developers, or the companies behind them had ceased operations altogether.}
Edmodo\footnote{https://en.wikipedia.org/wiki/Edmodo}, for example, was listed by three participants and only one discontinued using it, despite the company shutting down in 2022 
and discontinuing all support for the platform due to violating COPPA and being sued by the Federal Trade Commission.
Similarly, Blackboard~\cite{blackboard-phase-out} has phased out support for self-hosted deployments, potentially affecting institutions relying on that tool. 
In addition, the terms of service for ChatGPT prohibit use by individuals under the age of thirteen, making its presence in the K–12 sector a source of concern with regard to privacy and regulatory compliance.

\begin{SummaryBox}{Takeaway \#2:}
Some educators continue to use \utech even when it is legally challenged by government regulators or abandoned by the vendor.
\vspace{-2mm}
\end{SummaryBox}

\begin{table}[h]
    \small
    \renewcommand{\arraystretch}{1.2} 
    \caption{Most commonly listed unsanctioned applications with their primary category and frequency of educators. Here, ``Functionality'' means ``Functional convenience''.}
    \label{tab:top_apps}
    \centering
    \begin{tabular}{l c  c c}
        \toprule
        \textbf{Application} & \textbf{Category}& \textbf{\# K-12} & \textbf{\# HEI} \\
        \hline
        Kahoot! & Pedagogical & 42 & 18 \\
        ChatGPT & Functionality & 31 & 24 \\
        Google Classroom & Functionality & 43 & 12 \\
        Canva & Pedagogical & 33 & 14 \\   
        ClassDojo & Functionality & 31 & 5\\  
        Quizlet & Pedagogical & 23 & 11 \\
        Canvas & Functionality & 18 & 15 \\  
        Teachers Pay Teachers & Pedagogical & 22 & 3\\
        Youtube & Functionality &15 & 9\\ 
        Zoom & Functionality & 11 & 9\\
        Grammarly & Functionality & 9 & 11\\
        \bottomrule
    \end{tabular}
\end{table}

\subsubsection{Personal device use.} A large majority of our participants ($83\%$) reported using personal devices (e.g., laptop, tablet, smartphone) for teaching-related tasks. 
Among them, $93$ used personal devices daily, and $112$ used them several times per week. 
A total of $231$ participants reported downloading institutional documents---such as grade books or student records---onto personal devices. 
Of those, $127$ reported that these devices had automatic cloud backups enabled through personal accounts, potentially introducing further privacy and compliance risks.

\subsection{Discovery mechanisms}
\label{edtech_discovery_mechanism}

Past research has explored ubiquity, importance, as well as potential pitfalls~\cite{redmiles2016learned, redmiles2016think} in sourcing technology use and security advice from non-experts, including social, professional, and personal connections. 
To understand the primary unofficial sources for~\utech, we asked participants to indicate where they learned about each of the apps they listed.\footnote{We omitted similar questions for hardware tools since we only focused on personally owned devices}



\begin{table}[h]
    \small
    \renewcommand{\arraystretch}{1.2} 
    \caption{Prevalence of unsanctioned apps discovery mechanisms reported by educator participants}
    \label{tab:discovery_mechanisms}
    \centering
    \begin{tabular}{l r}
        \toprule
        \textbf{Discovery Mechanism} & \textbf{Count} \\
        \hline
        Professional Peers & 290 \\
        Admins or Supervisors & 141 \\
        Online Search or Browsing & 100 \\
        Advertisements or Sponsored Content & 100 \\   
        Friends or Family & 44 \\  
        Social Media or Influencers & 31 \\
        Workshops, Conferences, or Webinars & 29 \\     
        Previous Experience / Already Familiar & 4 \\
        \bottomrule
    \end{tabular}
\end{table}

Results are shown in Table~\ref{tab:discovery_mechanisms}: indeed, personal and professional connections play a dominant role in sourcing \utech. This ecosystem of trust---in colleagues, friends,  institutions, and public platforms---highlights the influence of informal channels in app discovery. The prevalence of peer recommendations suggests that educators frequently adopt unsanctioned tools based on advice from trusted professional networks, often bypassing formal vetting or approval processes.
These findings echo prior work on how individuals turn to familiar, non-expert sources for technology and security guidance~\cite{redmiles2016learned, redmiles2016think}, reinforcing the notion that trust-based, informal networks are a key driver in the adoption of~\utech.

\begin{SummaryBox}{Takeaway \#3:}
When it comes to the discovery of new \utech, most influences come from professional peers or supervisors.
\vspace{-2mm}
\end{SummaryBox}

\subsection{Selection and continued use}
\label{edtech_selection_criteria} 

This section presents the factors educators consider when adopting~\utech. Concretely, we asked participants to list 2--3 factors they considered for choosing each app, which we manually coded $17$ unique factors and grouped them into six primary categories (see Table~\ref{tab:app_factors_category}). 


\subsubsection{Reasons for selection.}
The most prevalent selection factor was \emph{pedagogical utility} ($n = 213$), including tools used for instruction, assessment, and content delivery—ranging from general education to subject-specific apps in math, reading, and science. 
\emph{Functional convenience} ($n = 155$) followed, capturing tools used for grading, lesson planning, communication, and learning management. 
Many tools spanned multiple subcategories; for example, Magic School AI\footnote{https://www.magicschool.ai/} was selected for its assessment and AI-driven personalization features.

Several apps under \emph{pedagogical utility} were selected for enabling \emph{gamified learning}, a trend particularly prominent among K-12 educators. 
Game-like elements were seen to enhance student engagement ($n = 30$), especially in supporting foundational skills such as reading and mathematics.
Twenty of the apps under \emph{functional convenience} were selected for AI features, including automated grading, content generation, and personalized learning plan development.
AI offers promising advancements in education, but it also presents risks like over-reliance, data privacy concerns, and potential biases~\cite{AI-education-risks}, requiring careful consideration during selection.

Beyond instructional use, educators prioritized \emph{accessibility} ($n = 187$) and \emph{cost} ($n = 164$), often preferring free, easy-to-access~\utech.
\emph{Familiarity} influenced tool selection for $n=117$ participants, often outweighing institutional alternatives
Educators cited long-term use, student comfort ``Zoom is more intuitive for students'' (P344), and frustration with frequent platform changes ``I wanted something more stable'' (P269). 
Some also noted that institutional tools were outdated or poorly maintained (P11).

Despite the diversity of motivations, \emph{security and privacy} were rarely cited.
\emph{Only 24 out of 375} participants explicitly mentioned ``security,'' ``privacy,'' ``secure,'' ``safe,'' or ``safety'' as a selection criterion, indicating that concerns about data protection were largely absent from most educators' EdTech-related decision-making processes.

\begin{SummaryBox}{Takeaway \#4:}
Familiarity often influenced \utech selection over institutionally-approved alternatives and far eclipsed security and privacy considerations.
\vspace{-2mm}
\end{SummaryBox}



\subsubsection{Reasons for continued use.}
After initial adoption, educators often made a deliberate choice to continue using~\utech. We asked them to state reasons for continuations and manually reviewed their responses to identify patterns. \emph{Ease of use} was the most commonly cited reason ($n = 294$), followed by \emph{improved student engagement} ($n = 108$), which was notedby $65\%$ of K-12 educators.

In HEI contexts, continued use was often driven by the need for specialized functionality not supported by institutionally acquired platforms ($n=61$).
Educators cited needs ranging from ``basic image editing'' (P300), ``reference apps for specific films and developer combinations'' (P151), to ``help with research papers'' (P52), or ``coding observational data'' in lab environments (P37).
Despite the widespread use of~\utech among K-12 and HEI educators, practical needs drive adoption, and security and privacy considerations are often left out---raising important questions about compliance, institutional oversight, and risk awareness.

\begin{table}[ht]
\caption{App adoption categories with frequency. 
}
\label{tab:app_factors_category}

\centering
\small
\begin{tabular}{l l r}
\toprule
\textbf{Category} & \textbf{Subcategories} & \textbf{\# Apps} \\
\hline
\multirow{10}{*}{\centering Pedagogical Utility}
 & Teaching Aides & 52 \\
 & General Ed & 31 \\
  & Games/Gamified Learning  & 30 \\
  & Math & 22 \\
  & Reading & 21 \\
   & English & 14 \\
   & Engineering & 13 \\
 & Assessments & 10 \\
 & Science & 10 \\
 & Special Education & 10 \\
\hline
\multirow{6}{*}{Functional Convenience}
 & Learning and Research Support & 74 \\
 & Learning Management & 43 \\
  & AI & 20 \\
 & Specialized Items & 9 \\
 & Editors & 9 \\
\hline
\multirow{1}{*}{\centering Accessibility}
 & -- & 187 \\
 
\hline
 \multirow{1}{*}{\centering Cost}
 & -- & 164 \\
 \hline
\multirow{2}{*}{\centering Familiarity}

 & Frustration with institutional tools & 84 \\
 & Long-term use, student comfort & 33\\

\hline
\multirow{1}{*}{\centering Security and Privacy}
 & -- & 19 \\
\bottomrule
\end{tabular}
\end{table}






%% file: 4-admin-survey.tex
\section{Security and Privacy (S\&P) Implications of Unsanctioned Technology Use (RQ2)}
\label{SP-implications}
Having examined the broad adoption of~\utech~and the predominance of practical considerations in tool selection, we now turn to the overlooked dimensions of security, privacy, and compliance to understand the risks associated with such~\utech~use.
We elicited educators' experience of invasive behaviors by apps they used in the past and their observations of any such behaviors by the apps listed and used currently. 
We similarly asked administrators to share their experience as well as risk perceptions from \utech~use. 
Finally, we report the implications of app integration, personal device use, and potential misunderstandings of regulations.




\subsection{S\&P risks observed by educators}
\label{observed_risks}
\subsubsection{Reasons for uninstallation of applications.} To find out what has made educators stop using technology, we asked educators if they have ever stopped using an application and why.
In total, 167 participants--108 being K-12---indicated that they encountered privacy or security-invasive behaviors that eventually led to the uninstallation of apps. We manually coded their experiences and identified 13 themes.  Table~\ref{tab:Reasons} lists the themes mentioned by at least two educators. The top reasons were related to excessive data collection, followed by sharing data with others and using the data to track users. Reasons stated by a single participant included apps being perceived as ``crypto miner'' (P13) or ``malware'' (P265).



\begin{table}[h]
    \renewcommand{\arraystretch}{1.2} 
    \caption{Educators' security-related reasons for uninstalling \utech.}
    \label{tab:Reasons}
    \centering
    \begin{tabular}{l r}
        \toprule
        \textbf{Reason} & \textbf{Count} \\
        \hline
        The app asked for unusual permissions &  52 \\
        The app collected more data than stated & 25 \\
        Shared data with third parties & 17 \\
        Tracked user & 15 \\   
        Experienced a breach & 11 \\  
        Sold data & 10 \\
        Asked for student information & 6 \\  
        Heard about issues second-hand & 5\\
        Administrators warned about security issues & 2\\
        \bottomrule
    \end{tabular}
\end{table}


Educators also shared their observations of behaviors, which they considered as privacy-violating, by apps they listed (\S~\ref{edtech_use_prevalence}). Forty-seven participants complained about apps because they thought some of the data the apps collected violated users' privacy. Several apps were mentioned by multiple participants; e.g., four participants complained about ChatGPT, three participants mentioned Blackboard,  and YouTube and Canvas had two complaints. Additionally, 62 participants expressed concerns about user data being sold or shared with third parties by several applications: five participants mentioned ChatGPT, four mentioned YouTube, and then two participants mentioned Canvas, Class Dojo, and Quizlet. 

    


\subsubsection{Leading to distrust in applications.} 
Many participants expressed disbelief in the technical ability of apps to safeguard user data. Thirty-eight participants explicitly stated that they distrusted at least one of their applications, and one participant distrusted all of their listed apps. Thirty-one participants believed that at least one app's developers were not competent to protect user privacy, with one participant holding this belief across all tools used.

Despite these concerns, all 62 participants continued using the same tools, citing reasons like the apps being deeply embedded in their workflow or not having suitable alternatives. Among them, ten participants had students register with their applications; three of them also stated that those applications were not FERPA compliant. These findings highlight a tension between perceived risk and pedagogical or practical utility that may outweigh data protection concerns in day-to-day decision-making.

\begin{SummaryBox}{Takeaway \#5:}
Educators overwhelmingly continue to use \utech despite their own concerns about user data safeguarding.
\vspace{-2mm}
\end{SummaryBox}
 
Participants who discontinued or uninstalled apps (for whatever reason) were asked follow-up questions about deleting data the apps collected. Only 13 participants indicated that the app they discontinued had an option to request data deletion, the rest were unsure about such a feature. Nine participants used the data deletion feature while one participant could not remember. The remaining three participants said they did not use that feature; notably, all three of them required students to register for that application. 

\begin{SummaryBox}{Takeaway \#6:}
Educators are mostly unaware of data deletion upon terminating use of \utech and, when they are aware, do not always use this feature.
\vspace{-2mm}
\end{SummaryBox}



\subsection{S\&P risks observed by IT administrators assisting educators} 
\label{sec:it_observed}
\subsubsection{Privacy incidents}
To complement educators’ perspectives, we surveyed administrators to understand institutional-level risks and incidents associated with~\utech~use.
Administrators shared several incidents they had to manage as a direct result of unsanctioned app use by educators through a free-response question in our admin survey. 
A2 noted, ``We have had third-party integrations scrape user data without being vetted and without contractual controls in place.'' 
While these cases were typically limited to small groups, such as a single class, the data exposed included names, emails, and, on occasion, student IDs. As the admin put it, this was ``information commonly available to instructors,'' but the access by third-party tools raised concerns.

Another issue was faculty routinely auto-forwarding university emails to other personally-owned email services. This became a major problem, as A9 noted ``None of our phishing controls or detection would detect anything after such emails had transited outside our managed environment, including phishing attack successes and account compromise.''

A more serious incident was described in connection with a breach involving the ticketing platform AudienceView mentioned by A17. They shared ``the product was purchased by an affiliate and did not go through our procurement process. The affiliate org handled the breach notifications.''


\subsubsection{Additional risks perceptions from administrators.}
In addition to past experiences with privacy incidents, in another free-form question, administrators shared potential threats from~\utech~use by educators that they were most concerned about. We coded the $21$ responses we received and identified concrete risks mentioned in them; we list them in Table~\ref{tab:risks_raised_by_admins} with the count of administrators mentioning them. Below we discuss the overarching themes and contextualize the risks using participant quotes.


\begin{table}[h]
    \small
    \centering
    \caption{Administrators' concerns regarding \utech~use}
    \begin{tabular}{ l r  }
    \toprule
    \textbf{Concerns} & \textbf{Count}  \\
    \midrule
    Data leakage & 13 \\
    Privacy invasion & 9\\
    Increased exposure of institutional ecosystem to threats & 6\\
    Increase regulatory or contractual liability & 5\\
    Increased operational overhead and liability & 5\\
    Lack of visibility & 4\\
    Malware, backdoors, apps with stealthy behavior & 4\\
    Lack of control over the system or data & 3\\
    Lack of IT/vendor support and security patches & 3\\
    (Rouge) AI tools use & 2\\
    Intellectual property theft & 2\\
    Data loss and corruption & 1\\
    Performance degradation & 1\\
    Loss of reputation & 1\\
    Misplaced trust & 1\\
    Silo effect  & 1\\
    \bottomrule
    \end{tabular}
    \label{tab:risks_raised_by_admins}
\end{table}

\paragraph{Threat to private data and increased liabilities.}
Administrators were most concerned about inadvertent leaks of PII or deliberate violations of students' privacy that can lead to subsequent misuse of private data. These risks may arise in different stages, as A2 explained, ``Inadvertent reading, storing and displaying of information...especially where it is held by a third party with no legal agreement or controls in place... where the third party is breached, or uses such information for other purposes (selling data, marketing, use to improve their product, etc).''

Several participants stated that such access to institutional data is a clear violation of data protection laws and institutional policies. A19 emphasized, ``Disclosure of covered information to third parties not under contract...this can commonly meet the criteria for a FERPA breach exposing the institution to serious liabilities.'' Beyond federal regulations, educators' use of apps can lead to financial obligations for the institute:
A19 added, ``There can be contractual liabilities where now you've obligated the organization to acquire the license due to your use.'' 



\smallskip
\paragraph{Increased exposure to invisible threats.}
Participants worried that the use of~\utech~create new risks because they can be faulty or deliberately malicious, stealing personal data or intellectual property. Worse, admins do not have any control over their behaviors or visibility into those risks, as A2 highlighted ``No visibility into risk introduced... No information of [the] security posture of the technology.'' Another participant, A11, added that these tools ``strip the controls [the institution] very intentionally design, build in and manage.''

Such risks were heavily discussed in the context of third-party ``add-ons'' and AI-based plugins that can be easily ``added'' into other apps (perhaps more trusted). A15 referenced ``insidious and rogue AI from third-party vendors in Zoom,'' and A16 highlighted the blind spots introduced by unvetted software: ``Not knowing what the application is doing and what backend communications it may establish.'' Similarly, A17 warned that ``The plug-in/software could be compromised to purposefully steal student data.'' Additionally, students may misplace their trust in these apps thinking that they were vetted or managed by their school.



\begin{SummaryBox}{Takeaway \#7:}
IT Administrators are concerned about introducing invisible S\&P vulnerabilities and non-compliance issues due to \utech{} use.
\vspace{-2mm}
\end{SummaryBox}

\subsection{Risks for regulatory non-compliance.}\label{sec:privacy:compliance}


In the US, FERPA~\cite{ferpa-faq} and HIPPA~\cite{hipaa} protect educational and health records, respectively; but they do not typically cover private entities unless they enter into contracts with education or health institutes~\cite{kelso_trust_2024}. Apps used in education or health contexts can proactively comply with these laws (e.g., Google Classroom and Zoom do that). We investigated the prevalence of such compliance and educators' knowledge in this matter.


We found that $107$ participants believed apps they listed were FERPA-compliant, $237$ believed at least one was compliant, and seven believed none were. 
Notably, $71$ participants were unsure whether any of their apps were FERPA-compliant, and $44$ participants ($21$ of whom were K–12 educators) explicitly stated that at least one app was not FERPA-compliant. For HIPAA, $81$ participants believed all apps were compliant, $182$ believed at least one was, and $10$ believed none were. Fifty-one participants were unsure about HIPAA compliance, and $55$ participants (including $23$ K–12 educators) stated that at least one app was not HIPAA-compliant. Uncertainty about legal compliance, however, did not result in the discontinuation of any app by any participants.

\begin{SummaryBox}{Takeaway \#8:}
Educators overwhelmingly continue to use \utech even when they are aware that the \utech is not compliant with legal regulations.
\vspace{-2mm}
\end{SummaryBox}


Since applicability and compliance information are not always readily apparent, to gain more clarity in this matter, we manually reviewed the privacy policy and terms of service of the top $15$ apps used by $52\%$ of the surveyed educators. Among these, $8$ indicated that they comply with FERPA (Blooket, Canva, Kahoot!, EdPuzzle, IXL, Prodigy, Zip Grade, ABC Mouse). Five apps---Quizlet, Khan Academy, Quizizz, Seesaw, and Padlet---were FERPA-compliant only when licensed institutionally. Only Teachers Pay Teachers did not comply with FERPA. Three apps: Teachers Pay Teachers, IXL, and Padlet, explicitly mentioned surveilling or tracking their users. 
COPPA most likely applies to each of these applications as students in K-12 were being registered for them. However, Khan Academy and Padlet stated that they only follow COPPA for verified student accounts. 

These findings shed light on additional issues due to apps not being institutionally sanctioned, as that would require them to comply with privacy laws and may additionally restrict data collection and use~\cite{kelso_trust_2024}.

\begin{SummaryBox}{Takeaway \#9:}
Almost half of the most-used \utech applications \emph{do not} comply with FERPA and/or COPPA when used as \utech (although they \emph{are} compliant when licensed institutionally).
\vspace{-2mm}
\end{SummaryBox}

\subsection{Additional risks from app integration and personal device use}  

Integrating unsanctioned apps into the institutional technology ecosystem was ubiquitous: in our sample, K-12 educators integrated 341 applications into their systems, while HEI educators implemented 116. Such integrations not only create blind spots, as mentioned by administrators, but also introduce new vulnerabilities to the the whole system~\cite{kelso_trust_2024}. Additionally, $338$ apps from K-12 and $140$ apps from HEIs required students to register using their institutional credentials such as email address. Students' personal data and education records can also be leaked through personal device use, as it is often associated with accessing or downloading educational documents. Having a personal cloud backup exacerbates this problem, as all that information can be automatically uploaded to the cloud server. More concerningly, other apps may have access to common directories, and apps with malicious intent can further disseminate this information. 

\begin{SummaryBox}{Takeaway \#10:}
Some \utech applications integrate with the institutional technology stack and require students to use their institutional credentials, introducing additional security risks.
\vspace{-2mm}
\end{SummaryBox} 


    

\section{Institutional and Personal Factors Influencing Unsanctioned Technology Use (RQ3)}
\label{RQ3}
This section presents the factors behind \utech{} use despite the associated risks identified above.

\subsection{Institutional policy: Existence, awareness, and enforcement}
\label{institute_policy_edtech_use}
All but one administrator reported that their institutions have general internal privacy policies; typically they extend beyond federal and state regulations by requiring additional audits for licensed applications~\cite{kelso_trust_2024}. However, knowledge of these policies among the educators was not widespread: only $30.3\%$ ($n = 30$) of K–12 and $24.8\%$ ($n = 80$) of HEI educators were aware that their institution had a policy on \utech use. Even among those who were aware, $22$ admitted to using unsanctioned apps in violation of policy. As P10 noted, ``[My institution doesn't] provide a workable path to utilize new and emerging technologies, so [I] have to go against it.'' 
Similarly, P68 stated, ``We are not supposed to use any application that is [not] already approved, but many, like myself, ignore the rule.'' Such violations were not a secret; many administrators stated that instructors often bring \emph{preferred} tools into the classroom without consulting institutional policy.  As A19 described, ``Almost at an individual level, everyone has their `solution' to teaching/learning `better' and brings it into the classroom with disregard.''

Administrators expressed frustration with the lack of policy enforcement. While institutions with policies provided online resources for faculty, actual enforcement was inconsistent---``There is a policy, but it needs enforcement'' (A8). 
A21 shared that policies were ``regularly ignored and rarely enforced.'' 
Some institutions had not updated their policies in five or more years, and one admin participant noted that their institution had not updated its policy in over a decade. 
Four admins were unsure whether a formal policy about unsanctioned EdTech use \emph{even existed}, and A20 plainly stated, ``There is no written policy.'' 
This security posture has allowed educators to use unsanctioned technologies without the fear of being punished by their administration.
A total of $107$ educator participants reported receiving institutional warnings about risks associated with unsanctioned apps. Of those, $33$ said they modified their behavior in response. Common adjustments included discontinuing the use of specific apps, switching to perceived more secure alternatives, enabling two-factor authentication, and limiting what data they shared with third-party platforms.

\begin{SummaryBox}{Takeaway \#11:}
Only a small minority of educators report modifying their behavior after receiving institutional warnings about their use of \utech.
\vspace{-2mm}
\end{SummaryBox}


\subsection{Institutional culture}
\label{institute_culture_edtech_use}
Administrators were generally aware that instructors used unsanctioned tools, and in several cases, they reported recommending such tools themselves. At the same time, administrators consider resource constraints and governance gaps as barriers to proactive oversight on~\utech~use.

\subsubsection{Administrators' awareness and informal endorsement.}
Seven admin participants acknowledged having advised educators to use unsanctioned applications, often to ``meet niche demands that our institutional platforms cannot manage, or where instructors are looking for free alternatives for their use or their students’ use'' (A4). Commonly recommended tools included ``Canva, Slack, and ChatGPT'' (A7). This can encourage educators to look for new technologies, and as seen in ~\ref{edtech_discovery_mechanism}, educators will listen to the administrator's recommendations.
Requests for formal integration of these tools into the institutional ecosystem were also widespread. 
All but three administrators in our study had received instructor requests to integrate unsanctioned apps into licensed platforms. 
For instance, A2 noted, ``We have instructors wishing to integrate many tools such as polling, scheduling, [and] citation software for use in classes.''
While a few administrators reported rejecting such requests outright, most indicated variability in decision-making, for instance, two said integration requests were never accepted, while others said they were accepted ``sometimes'' or ``most of the time.''
Twelve participants reported such requests triggered IT audits, while others were unsure whether audits were conducted.

\subsubsection{Resource constraints and governance gaps.}
When asked about proactive strategies to reduce risk, most administrators referenced existing guidelines or resources available to educators.
However, several barriers to implementation have been cited--primarily--cost, understaffing, and limited IT capacity. 
Three participants explicitly identified resource constraints as the main obstacle to proactive oversight of unsanctioned tools.
Some participants framed the issue not only as a matter of compliance or risk but as a structural governance failure. 
A14 noted: ``I don’t think this is a security and privacy issue---it’s a governance issue. Institutional leaders need to give clear guidance on when non-institutionally licensed tools can be used by instructors, who can make the decision about data used in those tools, and who can process exceptions. This isn’t an IT issue so much as it is a governance issue.''

\begin{SummaryBox}{Takeaway \#12:}
Permissive culture and approval from administrators are among the major drivers of \utech adoption.
\vspace{-2mm}
\end{SummaryBox}

\subsection{Educators' S\&P attitudes} 
\label{educator_sp_attitudes}
\subsubsection{Secure behavior.}
Here, we report educators' general attitudes and behavior about privacy and security and whether they are reflected in \utech~use. For the items measuring security practices~\cite{Egelman2015ScalingSeBIS}, $96.8\%$ of participants scored $3.5$ or higher on a 5-point scale, indicating generally high self-reported privacy and security awareness. There was minimal variation between HEI and K–12 participants: $90.3\%$ of HEI educators scored $3.5$ or higher, and $100\%$ of K–12 participants did as well. Educators in our sample demonstrated more secure behavior compared to the general public in the original study, as Table~\ref{tab:sp} shows. Additionally, there was no difference in the mean scores between those who previously stopped using invasive apps ($p>0.05$ for all items). Thus, following general security advice may not correlate to unsanctioned app use.   



\begin{table}[h]
	\renewcommand{\arraystretch}{1} 
	\caption{Security Behavior Intent scores---Mean (and Standard Deviation) on a 5-point scale---for security behaviors in our sample (left) compared to general population results reported by Egelman~\cite{Egelman2015ScalingSeBIS} (right). Bold text represents significant differences (all $p<.001$).}
	\label{tab:sp}
        \centering
	\begin{tabular}{p{4.7cm} r r}
		\hline
		 \textbf{Item} & \textbf{$\mu$ ($\sigma$)} & \textbf{$\mu$ ($\sigma$)~\cite{Egelman2015ScalingSeBIS}}  \\
		\hline
 I use a password/passcode to unlock my laptop or tablet & \textbf{4.82 (0.69)} & 3.78 (1.53)  \\  
 If I discover a security problem, I continue what I was doing because I assume someone else will fix it & 4.08  (1.07) & 4.08 (0.98) \\  
 I submit information to websites without first verifying that it will be sent securely & 3.75 (1.25) & 3.69 (1.08) \\  
 I try to make sure that the programs I use in personal devices are up-to-date & \textbf{4.38 (0.81)} & 3.78 (0.89)\\  
         \hline
	\end{tabular}
\end{table}

\subsubsection{Attitude toward other people's privacy.}
Since \utech{} use mostly risks the privacy of other people (i.e., students as seen in ~\ref{SP-implications}), we used three items from the VOPP scale~\cite{hasan_psychometric_2023} to measure how much educators value other people's privacy. Since did not use the full scale, we computed the reliability score for the selected subscale and found Cronbach’s $\alpha=0.91$, indicating high reliability~\cite{hasan_psychometric_2023}). Our sample mean was $4.57$ ($SD= 0.61$), suggesting educators' strong concern for others' privacy. Again, there was no difference between the groups that deleted invasive apps and those that did not. These concerns for other's privacy, however, do not stop educators from making students use \utech.

\subsubsection{Privacy prioritization}
Participants perceived attitude toward other's privacy, does not contextualize as making educators focus on privacy when selecting unsanctioned technology. 
Only one-third of participants ($n = 176$) listed privacy or security as one of the app selection criteria. When asked if they considered these aspects at all (in a separate question), only $72$ participants indicated that they considered privacy and security aspects for all apps they used, while this was true for at least one app for $218$ participants. 

\subsubsection{Educators' privacy literacy}
Privacy knowledge has been shown to impact behaviors~\cite{benamatiEmpiricalTestAPCO2017, mvungiAssociationsPrivacyRisk}. In our context, we measure it by asking participants if they knew about personally identifying information (PII), data brokers, and online tracking---concepts that are commonly associated with digital technology and data privacy. Two of the authors coded the free-form responses to these questions. We found that about half of the participants ($n=229$) defined all three terms correctly, while 104 and 32 answered two and one, respectively, questions, and nine participants failed to answer any of the questions. Looking at the individual responses, we find that 43 participants misunderstood. Some defined it inaccurately, such as ``[it's] to help stop people from intruding in others' privacy'' (P126), or described it in terms more aligned with classroom management: ``Tracking kids' data online. It is a great tool'' (P365).
Confusion extended to the term ``data brokers,'' which was misunderstood by $119$ participants ($98$ from K–12). 
One participant defined it as ``people who are paid to protect data'' (P123). Likewise, $35$ participants ($29$ from K–12) did not fully understand the term ``personally identifiable information.''  

We noticed, however, that educators who uninstalled privacy-invasive apps (\S~\ref{observed_risks}) performed better than others: more than 93\% of the 167 participants correctly answered questions about PII and online tracking, while 87\% correctly understood what is a data broker. We hypothesize that privacy literacy better enabled them to spot invasive behaviors compared to those who lack such knowledge. 

\begin{SummaryBox}{Takeaway \#13:}
Correct understanding of online privacy concepts among educators is correlated with higher uninstallation rates of privacy-invasive apps, hinting towards at education-based mitigations.
\vspace{-2mm}
\end{SummaryBox}



%% file: 5-discussion.tex
\section{Discussion}

\subsection{Implications of our findings}



\subsubsection{Implications for Educators.} 
We surfaced many privacy and security incidents experienced by educators as a side effect of using \utech, and yet they continued to use \utech~even when institutional alternatives were available, mainly due to habituation. The prominent role of habituation was observed among corporate employees as well~\cite{acken_who_2024}, which, in our case, even led to the use of apps and services that were banned after being sued by the United States Federal Trade Commission (FTC) for privacy violations~\cite{FTC-Edmodo} and apps that were labeled as ``critically risky'' by external evaluators (e.g., Quizlet~\cite{AppMicro}). Another key reason might be a lack of knowledge of privacy and safety risks from service providers (or application developers)~\cite{hielscherEmployeesWhoDont-2023}. Many participants did not know about tracking and data collection and selling by service providers, while they (self-reported) took actions such as using passwords or updating software. However, participants who knew about online tracking and data marketplace were more vigilant: they could detect privacy-invasive behaviors and stopped using apps that demonstrated those behaviors. Our participants also scored higher in how much they value other people's privacy compared to the general public; thus, their use of \utech{} that may violate others' privacy may seem paradoxical. However, this may again stem from a knowledge gap--if the risks are unknown, the use of \utech{} is unlikely to be perceived as a threat to others’ privacy.


\subsubsection{Implications for Students.}
Most of the apps and services we identified were ``free''; however, they likely rely on advertisements for revenue generation. Personalized advertisements encourage (more) private data collection, and ads might expose students to inappropriate content, or worse, scams~\cite{shaoUnderstandingInAppAds-2018}.
Beyond such immediate privacy and safety threats, we want to highlight some longer-term, but perhaps more damaging impacts on students: normalizing surveillance and devaluing students' agency. Students, especially younger ones who are more impressionable, may become normalized to being always surveilled, quantified, and compared to others, which can inhibit self-expressions and encourage self-censoring behaviors, negatively impacting their development~\cite{datafication-HE}. Prior studies reported students discomfort even with many uses of institutionalized technology~\cite{ multi-stakeholder-la, do-they-care, yang_discovering_2024, student-perception-privacy, students-privacy-perception-proctoring}; imposing more of the same might send a signal that their agency is not valued, and create privacy fatigue~\cite{privacy_fatigue_2017}. Moreover, big corporations, that provide critical services (such as insurance) and can be future employers, can obtain students' data through business agreements~\cite{zeide_learner_2018} or marketplaces~\cite{student-data-broker, marketplace-student-data, student-data-sold}, opening doors for exploitation and discrimination at a massive scale. Lastly, machine learning models can be used to leak (by prediction) further private information; e.g., interactions with digital tools have been shown to reveal demographics (e.g., gender and age~\cite{edtech-pets22}) and personality traits~\cite{peltonenPredictPersonalityMobileApp2020}, raising severe privacy, safety, and ethical concerns~\cite{kyritsi2019pursuit, edm_and_privacy}.


\subsubsection{Implications for Institutions.}
Beyond impacts on individuals, \utech~has privacy, security, and compliance implications for education institutes. 
Here, we discuss multiple main drivers of continued \utech{} use despite the associated risks. Past research highlighted the influence of workplace environment and peer behavior on encouraging learning and complying with cybersecurity measures~\cite{liInvestigatingImpactCybersecurity-2019}. 
Data from our study portray a permissive environment in which educators operate. Thus, in this lax environment, educators often do not bother to become aware of institutional policies that restrict \utech{} use, or simply ignore them. Administrators apparently lacked any real means to enforce those policies, while many encouraged \utech{} adoption. This creates a negative network effect as people prefer platforms that are popular among social and professional ties to privacy-friendly alternatives~\cite{zhangPrivacyVsConvenience-2024}. 

\subsubsection{Implications for Regulators.}
Applicable federal regulations, such as FERPA, are often criticized as not up to date for the increasingly digital environments~\cite{russell2018transparencystudentdata, zeide2015student}. Moreover, in case of violations of FERPA and HIPAA, institutes are held liable~\cite{ferpa-faq, hipaa-faq}, not individuals. It is also difficult, if not impossible, to link a data breach incident to an individual's (educator) use of an app. Thus, there is almost no personal risk for using \utech, but there are (perceived) immediate gains, such as improved productivity. As reported by Acken~\cite{acken_who_2024} in corporate settings, such a ``Performance-driven rule bending'' mentality is also common in the education context. Regulators might consider addressing this issue by setting baseline expectations for acceptable technology use, even in cases where formal approval processes are bypassed.

\subsubsection{Broader societal implications.}
The broader tech-ubiquitous environment of the education sector has experienced a dramatic rise in the use of digital technology in recent years, particularly fueled by the Covid-19 pandemic~\cite{datafication-HE, terasPostCovid19EducationEducation-2020, heads-in-the-clouds}. This may drive \utech~adoption even by people who are aware of the risks. This is because being surrounded by technology can create a feeling of powerless over personal information~\cite{hargittaiPrivacyApathy2016}, and people resign their privacy~\cite{resignation-app} and become ``privacy cynics'' where they abandon efforts and lose interest in privacy because the data will be out anyway~\cite{shalawadiDrConvenienceLove-PrivacyApathy2024}. Indeed, such coping mechanisms are not only common in the general population~\cite{sebergerStillCreepy2022, resignation-app, shalawadiDrConvenienceLove-PrivacyApathy2024, hargittaiPrivacyApathy2016}, but also among data stewards in organizations~\cite{popescuRolebasedPrivacyCynicism-2024}. This may explain why educators who were aware of the risks, as well as administrators, were accepting of \utech---if there is no escape, there is no point in rejecting the convenience \utech{} can provide. Moreover, when 96\% of school-approved apps send data to advertisers~\cite{apps-sell-data}, educators (and students) may feel demotivated in taking privacy-protective steps. 

\subsection{Recommendations.}

\subsubsection{Privacy-orientation and policy enforcement.} 
Awareness of privacy and safety risks from sensitive data collection by service providers was a key differentiator in behaviors, but such awareness among educators was also infrequent. Thus, besides security risks from external actors, cybersecurity awareness campaigns, training materials, and tutorials should also prioritize teaching privacy-related concepts and raise awareness about the additional risks that \utech creates. Often privacy and security mechanisms are perceived as barriers or disturbances to actual tasks~\cite{cram2023waste, sasseRebootingITSecurity-2023};
awareness or training campaigns can be contextualized to highlight how a privacy-respecting environment may foster freedom of expression and critical thinking among students which are among the key educational goals. A lack of awareness of institutional policy was too common, and their disregard was also notable; while in many cases it was because of convenience, in other cases, it was out of necessity. Thus, with stricter policy enforcement, it is also essential to work with educators to find an acceptable solution, such as providing a temporary, isolated environment to use a service, or helping them move to a more privacy-friendly alternative without disrupting their workflow~\cite{zhangPrivacyVsConvenience-2024}. 



\subsubsection{Cultivating a privacy-respecting environment.} To sustain privacy-protective behaviors, schools must create an environment where such behaviors are rewarded. A culture of open discussion about privacy and related concepts within organizations drives behavior change~\cite{chaudharyDrivingBehaviourChange-2024}. This is particularly important in this context as educators source \utech from peers and past research has shown that social and professional ties shape attitude and behaviors~\cite{zhangPrivacyVsConvenience-2024}. We recommend that privacy-conscious educators be assigned roles like ``privacy champions''~\cite{becker2017findingsecuritychampion}, who will transfer that knowledge to peers in casual settings and be rewarded for such efforts. Such local efforts have multiple benefits. Their knowledge is tuned to the local environment and more relevant to their peers. Unlike formal training by external experts, local learning sessions can be frequent, which can help replace past habits with new, privacy-protective behaviors~\cite{hielscher_taking_2022} and embed them into the workflow~\cite{sasseRebootingITSecurity-2023}. Local privacy champions can share their knowledge and recommendations in a casual setting that are more likely to be adhered to than formal warnings (which, our data suggests, get regularly ignored). Finally, these local efforts will over time diminish the negative network effect~\cite{zhangPrivacyVsConvenience-2024} and help establish a privacy-respecting environment.  

\subsubsection{Encourage critical evaluation of apps.} Equipped with privacy literacy and risk awareness, educators should be encouraged to evaluate apps before use. Administrators and peers knowledgeable in this matter can point to obvious indicators, such as asking for too many permissions or lacking a privacy policy. They can also identify and popularize external, more technically comprehensive assessments that are available online from authoritative organizations (e.g.,~\cite{AppMicro, scaryapps, ikeepsafe}).

\section{Conclusion}
Our study unveils the rampant use of unsanctioned technologies across K-12 and HEI institutions, despite institutional alternatives and institutional policies forbidding it (\S~\ref{institute_policy_edtech_use}). Our findings reveal that this landscape is not born out of malicious intent, but rather of necessity for educators to efficiently run their classrooms ($n = 213$)-with privacy not even a relevant factor for them ($n = 23$). Without any real consequences that fall on the educators, they turn to applications that make their lives easier with convenience at the forefront ($n = 155$). To address these challenges, educational institutions must raise awareness of the risks associated with unsanctioned technologies and foster a privacy-respecting environment that incentivizes the use of approved tools. Crucially, collaborating with educators---not simply enforcing compliance---will be essential in building a safer and more sustainable digital learning ecosystem for both teachers and students.


%% file: 7-appendix.tex
\section*{Appendix}
\appendix

\section{Why Sequential Survey Design?}
\label{why_sequential_design}
Sequential survey designs are best suited when insights from one survey inform the development of the other survey to gather rich data points from multiple perspectives~\cite{creswell2017designing}. 
K-12 and HEI instructors are the perfect stakeholder to learn about~\utech~as they engage directly with educational technologies on a day-to-day basis, navigate institutional policies firsthand, and depend on digital tools for teaching, research, and grading-related tasks.
In our context, educators are the \textit{de facto} acquirers of~\utech~while system administrators navigate the challenges posed by these apps to safeguard institute's infrastructure and ensure compliance with IT policies. 
Thus, we resort to a sequential survey design to capture both perspectives and utilize the insights from educators to infer questions for administrators to bring out a comprehensive understanding.

\section{Educator Survey Questionnaire}
\label{educator_questionnaire}
\noindent\textit{Instructions to Educators. } Please read the following instruction carefully. For this study, "not institutionally licensed or procured" refer to applications (for computers or mobile devices) that were not provided by your institution, rather instructors personally acquired them (including free apps and services). Similarly, "personal device" refers to devices owned by instructors and not provided by your institution. Our study purpose is to understand the use of such applications or devices for teaching and/or grading-related tasks from different viewpoints.

All questions were compulsory, if not marked optional.
\begin{itemize}
  \item If you agree to participate, check all the boxes below.
    \begin{itemize}
        \item I am age 18 or older,I have read this consent form or had it read to me
        \item I voluntarily agree to participate in this research study and I want to continue to the survey
        \item I have experience in using personal devices or personally-obtained software applications for teaching and research activities
    \end{itemize}
  \item Please enter your Prolific Id
  \item Please select the type of the institute you teach 
    \begin{itemize}
        \item K-12
        \item HEI
    \end{itemize}
  \item Please list 3-5 personal applications you have used for teaching, grading, or research-related tasks \\
  FOR EACH APP LISTED
    \begin{itemize}
        \item Why did you choose [APP]?
        \item List 2-3 most important factors you considered when selecting [APP].
        \item How did you learn about [APP]?
        \item Did you integrate [APP] to other institutionally licensed tools (such as Canvas or Zoom)? YES/NO/MAYBE
        \item Did students have to register with [APP] using their institutional credentials (e.g., email or id)? YES/NO/MAYBE
        \item Was there an alternative to [APP] available within the university's tool profile? If yes, why did you choose not to use it?
        \item Is [APP] still in use? YES/NO
        \item (optional) Did you transfer the data that [APP] collected to another tool or to back up data? If yes, Please briefly explain the transfer process.
        \item (optional) Did [APP] provide an option to request deletion of data before discontinuing? YES/NO/UNSURE
        \item (optional) Did you use the data deletion feature for [APP]? YES/NO/UNSURE
        \item (optional) Why you did not use the data deletion feature for [APP]? YES/NO/UNSURE

        \medskip
        \item Did you consider any security or privacy concerns when choosing and using [APP]?
        \item  Have you reported or observed any security or privacy issues with [APP]?
        \item Does [APP] access any data that you consider may violate the privacy of the data subject/owner?	
        \item To your knowledge, does [APP] share any data it collects with advertisers or other third parties?
        
        \medskip
        \item Was [APP] compliant with FERPA (Family Educational Rights and Privacy Act)?
        \item Was [APP] compliant with HIPPA (Health Insurance Portability and Accountability Act)?

        \medskip
        \item Please rate the following statements regarding [APP] on a 5-point scale from Strongly agree to Strongly disagree: 
        \begin{itemize}
            \item I trust this tool/application to keep user data private.	
            \item This tool/application is competent in safeguarding personal data.
        \end{itemize}

    \end{itemize}
    \medskip
  PERSONAL DEVICE USE:
  \item Have you ever used your personal mobile phone or computer for teaching or grading purposes? If no, why not?
  \item How frequently do you use your personal device (e.g., laptop, tablet, smartphone) for teaching-related tasks?
  \item Have you ever downloaded any academic documents (e.g., a grade sheet) to your personal device? - Grade Sheets
  \item Have you ever downloaded any academic documents (e.g., a grade sheet) to your personal device? - Student's Work	
  \item Have you ever downloaded any academic documents (e.g., a grade sheet) to your personal device? - Student data	
  \item Have you ever downloaded any academic documents (e.g., a grade sheet) to your personal device? - Other Documents (please explain)
  \item Does that personal device have automatic cloud backup enabled through your personal account?

  \medskip 
  INSTITUTIONAL POLICY and GENERAL APP-RELATED CONCERNS
  \begin{itemize}
      \item Are you aware of any institutional policies regarding the use of non-institutionally acquired tools for educational purposes?
      \item What are the institutional policies you are aware of? Please provide descriptions (or links if available).
      \item Has your institution ever reported/warned about any risk (such as a data breach) related to the use of independently acquired apps or devices?
      \item Did you change a behavior based on it? If so, what was it?
      \item Did you ever refrain from installing an app due to unusual permissions?
      \item Did you ever refrain from installing an app because it asked for a large number of permissions compared to the features provided?
      \item Did you ever refrain from installing an app because it asked for a large number of permissions compared to the features provided?
      \item Did you ever uninstall an app, after you heard that it is privacy-intrusive? How did the app invade privacy?
  \end{itemize}

  \medskip
  DIGITAL PRIVACY LITERACY 
  \item In 1-2 sentences, explain your understanding of Online Tracking
  \item In 1-2 sentences, explain your understanding of Data Brokers
  \item In 1-2 sentences, explain your understanding of Personally Identifiable Information (PII)

  \medskip
  DIGITAL SECURITY AND PRIVACY BEHAVIOR\\
  Please rate the following statements regarding [APP] on a 5-point scale from Strongly agree to Strongly disagree: 
  \begin{itemize}
      \item I use a password/passcode to unlock my laptop or tablet.
      \item If I discover a security problem, I continue what I was doing because I assume someone else will fix it.
      \item I submit information to websites without first verifying that it will be sent securely.
      \item I try to make sure that the programs I use in personal devices are up-to-date.
      \item It is important for me to protect other people's privacy even when it is difficult to do so.	
      \item Other people's privacy is valuable to me.
      \item It is important to protect other people's privacy even if I need to invest time and effort to do it.
      \item I am concerned that mobile apps may use my personal information for other purposes without notifying me or getting my authorization.
      \item When I give personal information to use mobile apps, I am concerned that apps may use my information for other purposes.
      \item I am concerned that mobile apps may share my personal information with other entities without getting my authorization.
    
  \end{itemize}
  
  \medskip
  DEMOGRAPHIC QUESTIONS
  \item Please select your gender.
  \item Please select your highest level of education.
  \item Please enter your academic discipline (e.g., Computer Science, Sociology, Accounting)
  \item Please enter your years of experience as an educator/instructor.
  \item Do you have any other comments regarding this study? At the end of your answer append the following sentence: ``No more comments''
\end{itemize}

\section{Administrator Survey Questionnaire}
\label{admin_questionnaire}
\noindent\textit{Instructions to administrators. } Please read the following instruction carefully. For this study, "not institutionally licensed or procured" refer to applications (for computers or mobile devices) that were not provided by your institution, rather instructors personally acquired them (including free apps and services). Similarly, "personal device" refers to devices owned by instructors and not provided by your institution. Our study purpose is to understand the use of such applications or devices for teaching and/or grading-related tasks from different viewpoints.

\begin{itemize}
    \item Have you ever recommended an instructor to use a technology (software or cloud service) that was not institutionally purchased or licensed?
    \item What tool did you recommend and why?
    \item Are you aware of any instructor's request to connect or integrate a tool (not institutionally licensed) into other tools that were institutionally licensed (such as Canvas or Google Classroom)?
    \item Which tool(s)? If you do not remember the exact name, please enter the type (such as class management system, grading assistant, etc.) and any other information you can recall about the tools.
    \item How frequently are such requests accepted at your institute?
    \item Do such integrations need to go through any security audits?
    \item In case of security or privacy incident(s) arise due to the use/integration of personally acquired applications or devices, who handles them? - Myself, Other employees of my institution, Third party service providers
    \item In your opinion or experience, what can be the most troubling security/privacy issue resulting from unlicensed technology use?
    \item Are there any proactive measures to prevent such issues at your institute?
    \item What proactive security and privacy measures does your institute have?
    \item What are the primary barriers to having proactive measures?
    \item Are you aware of any security or data privacy incident caused by the use of applications or devices not procured by the school? Please explain the incident(s) and how they were handled.
    \item Does your institutional policy have any clear recommendation for integrating non-institutionally acquired tools with institutional tools? 1) Clearly allowed for any tools, 2) Clearly denied for all tools, 3) Unsure, and 4) Need to review on a case-by-case basis
    \item What are the ways instructors (can) learn about the institutional policies? - 1) Online resources, 2) They are sent over email, 3) Other (please explain)
    \item Does your institutional policy have any clear recommendation for actions such as downloading student data by instructors on personally owned devices? YES/NO, explain
    \item Please explain	When was the last time the institutional policy regarding technology use and data security/privacy was updated?
    
    \medskip
    DEMOGRAPHIC QUESTIONS
    \item Please select your gender.
    \item Please select your age group.	
    \item Please enter your job title. This is the final question, so enter ``I am done'' at the end of your response.
    \item Please select the type of the institute you work at.
\end{itemize}